\documentclass[twocolumn,english]{revtex4-1}
\usepackage{mathptmx}
\usepackage[T1]{fontenc}
\usepackage[latin9]{inputenc}
\usepackage{color}
\usepackage{babel}
\usepackage{mathrsfs}
\usepackage{mathtools}
\usepackage{amsmath}
\usepackage{amssymb}
\usepackage{stmaryrd}
\usepackage{graphicx}
\usepackage{wasysym}
\usepackage[unicode=true,pdfusetitle,
 bookmarks=true,bookmarksnumbered=false,bookmarksopen=false,
 breaklinks=false,pdfborder={0 0 1},backref=false,colorlinks=true]
 {hyperref}
\hypersetup{
 linkcolor=blue,urlcolor={blue},filecolor={blue},citecolor={blue}}

\makeatletter
\usepackage{babel}
\usepackage{amsopn}\usepackage{amsthm}\usepackage{amscd}\usepackage{amsfonts}\usepackage{chngcntr}

\usepackage{times}

\counterwithout{figure}{section}
\counterwithout{equation}{section}
\usepackage{changepage}

\usepackage{babel}

\counterwithout{figure}{section}
\counterwithout{equation}{section}
\usepackage{times}
\usepackage{bm}
\usepackage[caption=false]{subfig}
\usepackage{floatrow}

\renewcommand{\fnum@figure}{FIG. \thefigure}

\makeatother

\begin{document}
\global\long\def\jm{J_{m}}  \global\long\def\dg{\mathbf{F}}  \global\long\def\dgcomp#1{F_{#1}}  \global\long\def\piola{\mathbf{P}}  \global\long\def\refbody{\Omega_{0}}  \global\long\def\refbnd{\partial\refbody}  \global\long\def\bnd{\partial\Omega}  \global\long\def\rcg{\mathbf{C}}  \global\long\def\lcg{\mathbf{b}}  \global\long\def\rcgcomp#1{C_{#1}}  \global\long\def\cronck#1{\delta_{#1}}  \global\long\def\lcgcomp#1{b_{#1}}  \global\long\def\deformation{\boldsymbol{\chi}}  \global\long\def\dgt{\dg^{\mathrm{T}}}  \global\long\def\idgcomp#1{F_{#1}^{-1}}  \global\long\def\velocity{\mathbf{v}}  \global\long\def\accel{\mathbf{a}}  \global\long\def\vg{\mathbf{l}}  \global\long\def\idg{\dg^{-1}}  \global\long\def\cauchycomp#1{\sigma_{#1}}  \global\long\def\idgt{\dg^{\mathrm{-T}}}  \global\long\def\cauchy{\boldsymbol{\sigma}}  \global\long\def\normal{\mathbf{n}}  \global\long\def\normall{\mathbf{N}}  \global\long\def\traction{\mathbf{t}}  \global\long\def\tractionl{\mathbf{t}_{L}}  \global\long\def\ed{\mathbf{d}}  \global\long\def\edcomp#1{d_{#1}}  \global\long\def\edl{\mathbf{D}}  \global\long\def\edlcomp#1{D_{#1}}  \global\long\def\ef{\mathbf{e}}  \global\long\def\efcomp#1{e_{#1}}  \global\long\def\efl{\mathbf{E}}  \global\long\def\freech{q_{e}}  \global\long\def\surfacech{w_{e}}  \global\long\def\outer#1{#1^{\star}}  \global\long\def\perm{\epsilon_{0}}  \global\long\def\matper{\epsilon}  \global\long\def\jump#1{\llbracket#1\rrbracket}  \global\long\def\identity{\mathbf{I}}  \global\long\def\area{\mathrm{d}a}  \global\long\def\areal{\mathrm{d}A}  \global\long\def\refsys{\mathbf{X}}  \global\long\def\Grad{\nabla_{\refsys}}  \global\long\def\grad{\nabla}  \global\long\def\divg{\nabla\cdot}  \global\long\def\Div{\nabla_{\refsys}}  \global\long\def\derivative#1#2{\frac{\partial#1}{\partial#2}}  \global\long\def\aef{\Psi}  \global\long\def\dltendl{\edl\otimes\edl}  \global\long\def\tr#1{\mathrm{tr}#1}  \global\long\def\ii#1{I_{#1}}  \global\long\def\dh{\hat{D}}   \global\long\def\sp{\psi}     \global\long\def\vp{\boldsymbol{\phi}}        \global\long\def\pt{\mathscr{PT}}
 \global\long\def\inc#1{\dot{#1}}  \global\long\def\sys{\mathbf{x}}  \global\long\def\curl{\nabla}  \global\long\def\Curl{\nabla_{\refsys}}  \global\long\def\piolaincpush{\boldsymbol{\Sigma}}  \global\long\def\piolaincpushcomp#1{\Sigma_{#1}}  \global\long\def\edlincpush{\check{\mathbf{d}}}  \global\long\def\edlincpushcomp#1{\check{d}_{#1}}  \global\long\def\efincpush{\check{\mathbf{e}}}  \global\long\def\efincpushcomp#1{\check{e}_{#1}}  \global\long\def\elaspush{\mathbf{{C}}}  \global\long\def\elecpush{\boldsymbol{\mathcal{A}}}  \global\long\def\elaselecpush{\boldsymbol{\mathcal{B}}}  \global\long\def\disgrad{\mathbf{h}}  \global\long\def\disgradcomp#1{h_{#1}}  \global\long\def\trans#1{#1^{\mathrm{T}}}  \global\long\def\phase#1{#1^{\left(p\right)}}  \global\long\def\elecpushcomp#1{\mathcal{A}_{#1}}  \global\long\def\elaselecpushcomp#1{\mathcal{B}_{#1}}  \global\long\def\elaspushcomp#1{\mathbf{C}_{#1}}  \global\long\def\dnh{\aef_{DH}}  \global\long\def\dnhc{\mu\lambda^{2}}  \global\long\def\dnhcc{\frac{\mu}{\lambda^{2}}+\frac{1}{\matper}d_{2}^{2}}  \global\long\def\dnhb{\frac{1}{\matper}d_{2}}  \global\long\def\afreq{\omega}  \global\long\def\dispot{\phi}  \global\long\def\edpot{\varphi}  \global\long\def\afreqh{\hat{\afreq}}  \global\long\def\phasespeed{c}  \global\long\def\bulkspeed{c_{B}}  \global\long\def\speedh{\hat{c}}  \global\long\def\dhth{\dh_{th}}  \global\long\def\bulkspeedl{\bulkspeed_{\lambda}}  \global\long\def\khth{\hat{k}_{th}}  \global\long\def\p#1{#1^{\left(p\right)}}  \global\long\def\maxinccomp#1{\inc{\outer{\sigma}}_{#1}}  \global\long\def\maxcomp#1{\outer{\sigma}_{#1}}  \global\long\def\relper{\matper_{r}}  \global\long\def\sdh{\hat{d}}  \global\long\def\iee{\varphi}  \global\long\def\effectivemu{\tilde{\mu}}  \global\long\def\fb#1{#1^{\left(1\right)}}  \global\long\def\mt#1{#1^{\left(2\right)}}  \global\long\def\phs#1{#1^{\left(p\right)}}  \global\long\def\thc{h}  \global\long\def\state{\mathbf{s}}  \global\long\def\harmonicper{\breve{\matper}}  \global\long\def\kb{k_{B}}  \global\long\def\cb{\bar{c}}  \global\long\def\mb{\bar{\mu}}  \global\long\def\rb{\bar{\rho}}  \global\long\def\wavenumber{k}  \global\long\def\rh{\hat{\mathbf{r}}}  \global\long\def\zh{\hat{\mathbf{z}}}  \global\long\def\th{\hat{\mathbf{\theta}}}  \global\long\def\lz{\lambda_{z}}  \global\long\def\lt{\lambda_{\theta}}  \global\long\def\lr{\lambda_{r}}  \global\long\def\st{\Omega}  \global\long\def\stz{\Psi}  \global\long\def\ste{\varphi}  \global\long\def\stze{\phi}  \global\long\def\lap{\mathcal{M}}  \global\long\def\vh{\hat{V}}  \global\long\def\ch{\hat{c}}  \global\long\def\wh{\hat{\omega}}  \global\long\def\rb{\bar{r}}  \global\long\def\cthick{h}  \global\long\def\vth{\Delta\vh_{th}}  \global\long\def\kco{\kh_{co}}  \global\long\def\normv{\Delta\hat{V}}  \global\long\def\qh{\hat{q}_{A}}  \global\long\def\kh{\hat{k}}  \global\long\def\lzt{\tilde{\lambda}_{z}}  \global\long\def\cratio{\gamma}  \global\long\def\torusvar{\zeta}  \global\long\def\torusfun{\mathcal{T}}  \global\long\def\gapdensity{P}  \global\long\def\gapdom{\mathbb{D}}  \global\long\def\torus{\mathbb{T}}  \global\long\def\se{\Psi}  \global\long\def\slop{a} \global\long\def\disp{\mathbf{u}} \global\long\def\nh{\mathbf{n}} \global\long\def\mh{\mathbf{m}} \global\long\def\strain{\boldsymbol{\epsilon}}
\global\long\def\ux{U\left(x\right)}
\global\long\def\uxm#1{U_{#1}\left(x\right)}

\global\long\def\ux{U\left(x\right)}%
\global\long\def\phaseone#1{#1_{a}}%
\global\long\def\phasetwo#1{#1_{b}}%
\global\long\def\period{l}%
\global\long\def\scalaru{u}%
\global\long\def\iphase#1{#1_{i}}%
\global\long\def\impedancemis{\gamma}%
\global\long\def\uxm#1{U_{#1}\left(x\right)}%
\global\long\def\px{\Psi\left(x\right)}%
\global\long\def\pxm#1{\Psi_{#1}\left(x\right)}%
\global\long\def\pxmh#1{\hat{\Psi}_{#1}\left(x\right)}%
\global\long\def\sys{\mathbf{x}}%
\global\long\def\cp#1#2{\left(#1,#2\right)}%
\global\long\def\inner{\mathcal{I}}%
\global\long\def\cn{C_{n}}%
\global\long\def\dc{\Delta c}%
\global\long\def\hh{\hat{H}}%
\global\long\def\dka{\Delta k_{nm}}%
\global\long\def\lep{\alpha^{EP}}%
\global\long\def\hmatz{\mathsf{H}^{\left(0\right)}}%
\global\long\def\hmato{\mathsf{H}^{\left(1\right)}}%
\global\long\def\hp{\mathsf{H}}%
\global\long\def\newmacroname{}%
\global\long\def\ddx{\frac{\mathrm{d}\ \,}{\mathrm{d}x}}%
\global\long\def\ddxs{\frac{\mathrm{d}^{2}\ \,}{\mathrm{d}x^{2}}}%
\global\long\def\ddxa#1{\frac{\mathrm{d}#1}{\mathrm{d}x}}%
\global\long\def\ddxsa#1{\frac{\mathrm{d}^{2}#1}{\mathrm{d}x^{2}}}%
\global\long\def\pt{\mathcal{PT}}%
\global\long\def\cx{c\left(x\right)}%
\global\long\def\nx{n\left(x\right)}%
\global\long\def\mux{\mu\left(x\right)}%
\global\long\def\rx{\rho\left(x\right)}%
\global\long\def\parameter{\alpha}%
\global\long\def\innerc#1#2{\boldsymbol{(}#1,#2\boldsymbol{)}}%
\global\long\def\fn{\varsigma_{n}}%

\begin{abstract}
 Recent years have seen a fascinating pollination of ideas from quantum
theories to elastodynamics---a theory that phenomenologically describes
the time-dependent macroscopic response of materials. Here, we open
route to transfer additional tools from non-Hermitian quantum mechanics.
We begin by identifying the differences and similarities between the
one-dimensional elastodynamics equation and the time-independent Schrödinger
equation, and finding the condition under which the two are equivalent.
Subsequently, we demonstrate the application of the non-Hermitian
perturbation theory to determine the response of elastic systems;
calculation of leaky modes and energy decay rate in heterogenous solids
with open boundaries using a quantum mechanics approach; and construction
of degeneracies in the spectrum of these assemblies. The latter result
is of technological importance, as it introduces an approach to harness
extraordinary wave phenomena associated with non-Hermitian degeneracies
for practical devices, by designing them in simple elastic systems.
As an example of such application, we demonstrate how an assembly
of elastic slabs that is designed with two degenerate shear states
according to our scheme, can be used for mass sensing with enhanced
sensitivity by exploiting the unique topology near the exceptional
point of degeneracy. 
\end{abstract}
\title{Linking scalar elastodynamics and non-Hermitian quantum mechanics}
\author{Gal Shmuel$^{1}$ and Nimrod Moiseyev$^{2,3}$}
\affiliation{$^{1}$Faculty of Mechanical Engineering, Technion--Israel Institute
of Technology, Haifa 32000, Israel}
\affiliation{$^{2}$Faculty of Chemistry, Technion--Israel Institute of Technology,
Haifa 32000, Israel}
\affiliation{$^{3}$Faculty of Physics, Technion--Israel Institute of Technology,
Haifa 32000, Israel}
\maketitle

\section{Introduction}

The physics of matter at the subatomic level is described by quantum
mechanics. The computational complexity associated with the theory
at the macroscopic scale renders it infeasible to describe the observable
mechanics of materials, and hence continuum mechanics is used \citep{trusnoll04}.
Despite the huge difference in the length scale that the two theories
were developed for, fascinating realizations of quantum phenomena
were demonstrated using macroscopic systems in recent years \citep{Ma2019db}.
Examples include the Hall effect \citep{Miniaci2019prb,Lera2019prb},
geometric phase \citep{Xiao2015natphys}, and negative refraction
\citep{Srivastava2016jmps,Willis2016jmps,NEMATNASSER2019MOM,Hou2018PRApplied,Lustig2019}.
Special attention is given to extraordinary transport properties based
on $\pt$ symmetry \citep{ruter2010observation,Graefe2011pra,Fleury2015,Cummer2016,Christensen2016prl,Achilleos2017PRB,hou2018jap,Merkel2018prb},
which corresponds to the commutativity of an operator with combined
parity-time reversal operators. This concept originated form the discovery
in quantum mechanics that Hamiltonians exhibiting this symmetry can
have real eigenvalues, even if they are not Hermitian \citep{Bender1998PRL}.
One the advantages of Non-Hermitian Quantum Mechanics (NHQM) is its
quantification of the conditions for the existence of $\pt$-symmetric
Hamiltonians with a real spectrum \citep{bender2002prl}. 

The source of these analogies originates from the connection between
the governing equations in the different branches of physics. The
analogy between the time-independent Schrödinger equation and the
scalar elastodynamic equation that appears in part of the literature
identifies the transformation
\[
\hat{V}(x)-E\rightarrow\frac{\omega^{2}}{c^{2}(x)}
\]
where $\hat{V}(x)$ is the potential in the quantum Hamiltonian and
$E$ is the energy, and in the elastic counterpart $\omega$ and $c(x)$
are the wave frequency and  velocity, respectively. However, this
analogy is flawed, since it mixes the operator and its eigenvalues.
Using a simple transformation, we here first identify the term that
appears in the one-dimensional elastodynamics equation and absent
from the time-independent Schrödinger equation. In turn, this derivation
allows us to determine the condition under which the two equations
are equivalent. Following this analysis, we apply tools from the non-Hermitian
formalism of quantum mechanics to elastodynamics, in addition to those
transferred recently \citep{lu2018level,ZHANG2019jmps}, as described
next.

First, we show the application of the  time-independent Rayleigh-Schrödinger
perturbation theory of quantum mechanics to elastodynamics \citep{HIRSCHFELDER1964,fernandez2000introduction}.
This theory provides the solutions of a perturbed Hamiltonian in terms
of a series expansion about an Hermitian Hamiltonian, where its non-Hermitian
formalism determines the radius of convergence by extending the perturbation
to the complex plane. By way of example, we consider an elastic assembly
composed of a PMMA slab that is perfectly bonded between two steel
slabs which are fixed at the ends. We apply the aforementioned theory
to calculate the shear response of an elastic assembly whose properties
are complex perturbations of the original assembly. This response
is given in terms of a perturbation expansion, for which we calculate
its radius of convergence \citep{Certain1975}. Importantly, in this
process we also calculate the \emph{exceptional point} (EP)---the
point at which the spectrum of the perturbed assembly has a non-Hermitian
degeneracy, where two of its eigenmodes coalesce, together with their
corresponding complex frequencies \citep{Moiseyev1980PRA,Miri2019science,Ozdemir2019cr}.
This occurs in our example for an assembly comprising a lossy slab
with specific viscoelastic shear modulus \citep{Laude2013prb}.

Subsequently, we present a proof of concept how this assembly, i.e.,
an elastic assembly with non-Hermitian degeneracy in its spectrum,
can be utilized for mass sensing with enhanced sensitivity. Specifically,
by a combination of algebraic arguments and numerical calculations,
we show that when a mass is deposited, the degenerate frequency of
the elastic assembly splits into two frequencies, such that the splitting
is proportional to the square root of the mass. This phenomenon is
the physical manifestation of the topology near an EP in the spectrum
of our elastodynamic problem. Accordingly, measurement of the frequency
splitting quantifies the weight of the deposited mass, with higher
mass responsivity at small masses. By contrast, standard mechanical
sensors measure the shift in the mechanical resonant frequency, which
scales linearly with the deposited mass, hence of inferior sensitivity
at small masses \citep{Liu2013qf,He2015rp}. Indeed, the square-root
topology near an EP has been used in other systems for sensing \citep{Wiersig2014prl,Wiersig2016pra},
most recently by Djorwe et al.$\ $\citep{Djorwe2019prapplied} using
optomechanical cavities coupled by mechanical resonators.

The second analogy we draw is between the NHQM formalism of the particle
in a box model and the previous elastic assembly, when the steel slabs
are semi-infinite. This problem corresponds to a non-Hermitian Hamiltonian,
owing to (radiation) outgoing boundary conditions. Only in the non-Hermitian
formalism of quantum mechanics the poles of the scattering matrix
are associated with metastable states, where the imaginary part of
the poles provides the resonance width, or rate of decay of resonance
state \citep{moiseyev2011book}. Here, we obtain the physical counterparts
of these quantities in the elastodynamic settings. Specifically, we
show that the imaginary part of the poles of the elastic scattering
matrix equals half the decay rate of the mechanical energy in the
PMMA slab, associated with leaky modes in the elastic assembly. 

Finally, we develop a real perturbation theory for the non-Hermitian
system using the NHQM complex scaling method \citep{Moiseyev1979,MOISEYEV1998}.
With the framework developed in this Section, we are able to derive
the eigenstates of the perturbed Hamiltonian as an expansion about
a non-Hermitian Hamiltonian with real parameters, such as the stiffness
and length of a fourth slab in our example. Thereby, we constitute
a framework to analyze and construct degeneracies by real perturbations,
although its numerical study is beyond our scope here \footnote{Such a study will presumably require two-dimensional systems \citep{Lustig2019}.}.
It has already been established that systems exhibit extraordinary
behavior in the vicinity of non-Hermitian degeneracies, such as ultra-sensitivity
\citep{Zhong2019prl} (as we also demonstrate in the sequel), Berry
phase acquiring \citep{Mailybaev2005PRA}, and asymmetric scattering
properties \citep{Shen2018prmat,Thevamaran2019}. Accordingly, our
framework offers an approach to achieve such extraordinary wave phenomena
by designing non-Hermitian degeneracies in simple elastic systems,
without the need for external gain and loss as in the works mentioned
earlier. 

The results described above are presented in the following order.
Section \ref{sec:The-elastodynamics-equations} provides a short summary
of the elastodynamics equations, and specifically their reduced form
in the scalar (one-dimensional) setting. Section \ref{sec:Similarities-and-differences}
identifies the similarities and differences between the scalar elastodynamic
equation and the time-independent Schrödinger equation, based on a
transformation we develop. Section \ref{sec:EP} formulates the elastic
counterpart of the NHQM time-independent Rayleigh-Schrödinger theory,
demonstrates the calculation of the radius of convergence for our
model slabs problem, and importantly determines the EP in the spectrum
of the slabs. Section \ref{sec:Application-to-mass-sensing} demonstrates
how the elastic assembly that exhibits a non-Hermitian degeneracy
can be utilized for mass sensing with enhanced sensitivity. Section
\ref{sec:radiation} details the calculation of metastable states
and energy decay when the steel slabs are semi-infinite, using the
analogy with the NHQM particle in a box model. The development of
a real perturbation theory for non-Hermitian elastic systems is carried
out in Section \ref{sec:theory-for-real}. A summary of our results
and outlook concludes the paper in Section \ref{sec:Summary}. 


\section{\label{sec:The-elastodynamics-equations}The elastodynamics equations
in Continuum mechanics}

The continuum governing equations are based on the hypothesis that
the inter-particle forces can be replaced by the \emph{stress} tensor
field $\cauchy$; in terms of $\cauchy$, the balance of linear momentum
yields \citep{graff1975wave}

\begin{equation}
\grad\cdot\cauchy(\sys,t)=\rho\left(\sys,t\right)\ddot{\disp}(\sys,t),\label{eq:eom}
\end{equation}
where $\rho$ is the mass density and $\disp$ is the displacement
vector field of material points. The stress is related to the displacement
field via the \emph{constitutive} equation

\begin{equation}
\cauchy=\elaspush\left(\sys\right)\grad\disp,\label{eq:hooke}
\end{equation}
where $\elaspush$ is the fourth-order elasticity tensor. If the material
is locally isotropic, the tensor $\elaspush$ is a function of the
La\'{m}e parameters $\mu\left(\sys\right)$ and $\lambda\left(\sys\right)$,
and the combination of Eqs.$\ $\eqref{eq:eom}-\eqref{eq:hooke}
can be put in the form 
\begin{equation}
\left\{ \grad\left[\lambda\left(\sys\right)\grad\cdot+2\mu\left(\sys\right)\grad\cdot\right]-\grad\times\left[\mu\left(\sys\right)\grad\times\right]\right\} \disp=\rho\left(\sys\right)\ddot{\disp}.\label{eq:eom shear pressure}
\end{equation}
 Eq.$\ $\eqref{eq:eom shear pressure} exposes the unique coupling
in elastodynamics between the volumetric part of the vector field,
proportional to $\grad\cdot\disp$, and its transverse or shear part,
proportional to $\grad\times\disp$. This coupling has a significant
effect on the Hermiticity of the system, discussed elsewhere \citep{Lustig2019}.
When considering one-dimensional motions, the coupling is eliminated
and the problem reduces to a scalar one. Using the ansatz $u(x,t)=U(x)e^{-i\omega t}$,
Eq.$\ $\eqref{eq:eom shear pressure} then reduces to
\begin{equation}
-\ddx\tilde{\mu}\ddx U\left(x\right)=\rho\left(x\right)\omega^{2}U\left(x\right),\label{eq:1dhetro}
\end{equation}
where $\tilde{\mu}=\mu$ when the displacements are normal to the
$x$ direction (termed transverse or shear waves), and $\tilde{\mu}=\lambda+2\mu$
when the displacements are along the $x$ direction (termed pressure
or volumetric waves). In what follows we focus on the former, bearing
in mind that the same analysis holds for the latter, by carrying out
a change of modulus.

\section{\label{sec:Similarities-and-differences}Similarities and differences
between the 1D elastodynamic equation and the %
 time-independent Schrödinger equation}

The objective of this Section is to transform the equation of elastodynamics
in the one-dimensional case to a Schrödinger-type equation, in order
to highlight the similarities and differences between them. To this
end, we first multiply Eq.$\ $\eqref{eq:1dhetro} by $\mu^{-1}\left(x\right)$,
and define $c^{2}\left(x\right)=\mu\left(x\right)/\rho\left(x\right)$
to obtain
\begin{equation}
-\frac{1}{\mu}\ddx\mu\ddx U=\frac{\omega^{2}}{c^{2}}U.\label{eq:s1}
\end{equation}
Observe that in terms of the variable $\eta\left(x\right)=\mu^{-1}\left(x\right)$,
the left-hand side equals
\begin{equation}
-\eta\ddx\frac{1}{\eta}\ddx U=\left[\frac{1}{\eta}\ddxa{\eta}\ddx-\ddxs\right]U.\label{eq:s2}
\end{equation}
 By further defining $n^{2}\left(x\right)=\eta\left(x\right)$, we
rewrite Eq.$\ $\eqref{eq:s1} as
\begin{equation}
\left[2\frac{1}{n}\ddxa n\ddx-\ddxs\right]U=\frac{\omega^{2}}{c^{2}}U.\label{eq:motion 1d3-1}
\end{equation}
Finally, we employ the transformation $U\left(x\right)=n\left(x\right)\Psi\left(x\right)$
and multiply Eq.$\ $\eqref{eq:motion 1d3-1} by $c^{2}/n$ to achieve
the form
\begin{equation}
\hat{H}\Psi\left(x\right)=\omega{}^{2}\Psi\left(x\right),\label{eq:schrodinger}
\end{equation}
where $\omega^{2}$ is the eigenvalue, and $\hat{H}=\hat{T}+\hat{V}_{CM}$
with 
\begin{equation}
\hat{T}=-c^{2}\left(x\right)\ddxs,\quad\hat{V}_{CM}=2\left(\frac{c\left(x\right)}{n\left(x\right)}\ddxa n\right)^{2}-\frac{c^{2}\left(x\right)}{n\left(x\right)}\ddxsa n.\label{eq:TV}
\end{equation}
The operator $\hat{V}_{CM}$, which is a local function of $x$ and
does not involve spatial derivatives, can be interpreted as the potential
of a conservative force. By further separating $\hat{T}$ according
to 
\begin{equation}
\hat{T}=\hat{T}_{QM}+\hat{T}_{CM},\ \ \hat{T}_{QM}=-\ddx c^{2}\left(x\right)\ddx,\ \ \hat{T}_{CM}=\ddxa{c^{2}}\ddx,\label{eq:T}
\end{equation}
we can identify $\hat{T}_{QM}$ with the kinetic energy operator in
the Schrödinger equation of an electron  with an effective mass 
\begin{equation}
m_{\mathrm{eff}}(x)=\frac{1}{2c^{2}(x)},\label{eq:emass}
\end{equation}
that varies when the electron traverses different semiconductors.
The difference between the equations thus amounts to $\hat{T}_{CM}$---this
term does not have the form of a kinetic energy operator nor a potential,
as it involves one spatial derivative. 

\begin{figure}[t]
\centering\includegraphics[width=1\textwidth]{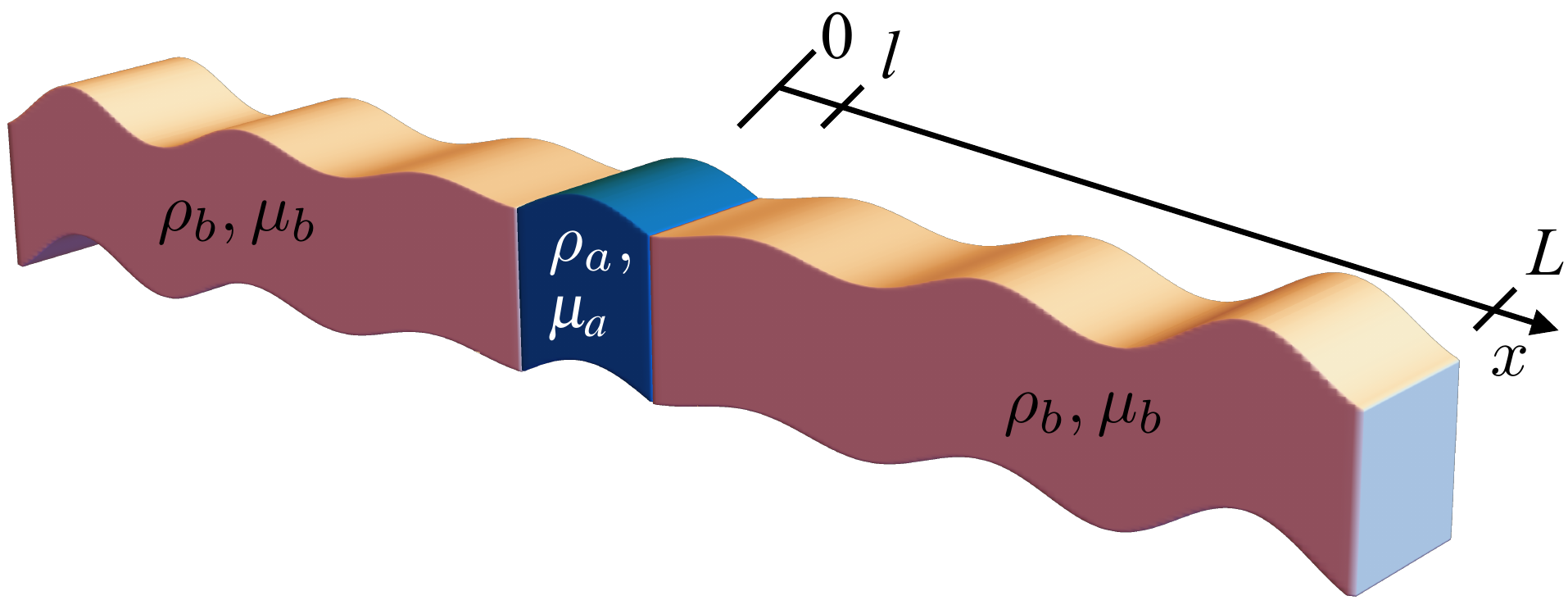}\caption{An assembly of three elastic slabs. The Hermitian case corresponds
to fixed boundaries at $x=\pm L$, and the non-Hermitian case corresponds
to an infinite assembly with outgoing boundary conditions. }

{\small{}{}\label{fig:slabs}}{\small\par}
\end{figure}

To draw the analogy with the NHQM model problem of a particle in a
box with outgoing boundary conditions, we consider the prevalent case
of a solid that is composed of different homogeneous slabs. The medium
properties are therefore piecewise constant. For simplicity, we consider
two constituents, say, material $a$ with $\rho_{a}$ and $\mu_{a}$,
which is perfectly bonded at $x=\pm l$ to two infinite slabs made
of a stiffer material $b$ with $\rho_{b}$ and $\mu_{b}>\mu_{a}$
(Fig.$\ $\ref{fig:slabs}).

\floatsetup[figure]{style=plain,subcapbesideposition=top}

If we further assume that 
\begin{equation}
\frac{\phaseone{\mu}}{\rho_{a}}=\frac{\mu_{b}}{\rho_{b}},\label{eq:velocitycontrast}
\end{equation}
then $\hat{T}_{CM}$ vanishes; in this case---and this case only---there
is an exact analogy between the one-dimensional equation of elastodynamics
and the time-independent Schrödinger equation. 
 The corresponding potential $\hat{V}_{CM}$ exhibits a potential
well between two infinite barriers (spikes), owing to the jump discontinuities
of $\mux$, and hence of $\nx$.  In the equivalent quantum particle
in a box model, there are resonance phenomena and metastable states
associated with complex eigenvalues that are elegantly analyzed and
explained using the NHQM formalism \citep{moiseyev2011book}. In
the sequel, we will show that this formalism establishes a powerful
machinery to study corresponding elastodynamic phenomena, even when
restriction \eqref{eq:velocitycontrast} is removed and the exact
analogy is broken. Before we proceed, it is useful to note that for
two-dimensional elastodynamics, by contrast, an exact analogy with
the time-independent Schrödinger equation always exists. To show this,
it is sufficient to consider anti-plane shear waves of the form 
\begin{equation}
u\left(x,y,t\right)=Y\left(y\right)e^{i\left(k_{x}x-\omega t\right)}\label{eq:uxy}
\end{equation}
propagating in a medium that is laminated in the $y$ direction. In
each lamina, Eq.$\ $\eqref{eq:eom shear pressure} reduces to the
following equation for $Y\left(y\right)$
\begin{equation}
\left[\frac{\mathrm{d}^{2}}{\mathrm{d}y^{2}}+\frac{\omega^{2}}{c^{2}\left(y\right)}\right]Y\left(y\right)=k_{x}^{2}Y\left(y\right).\label{eq:ev for Y}
\end{equation}
In this case, it is possible to identify $k_{x}^{2}$ (not $\omega^{2}$)
as the eigenvalue to be determined, $\mathrm{d}^{2}/\mathrm{d}y^{2}$
with the kinetic energy operator, and $\omega^{2}/c^{2}\left(y\right)$
with the potential. Physically, Eq.$\ $\eqref{eq:ev for Y} represents
the question: given excitation frequency and mechanical properties,
what would be the propagation constant in the direction perpendicular
to the material modulation?\emph{ cf.}$\ $Ref.$\ $\citep{pick2018pra}
on a similar analogy between the time-independent Schrödinger equation
and Maxwell equations. 

Returning to the one-dimensional problem, we consider next the transformation
$U\left(x\right)=\rho^{-1/2}\left(x\right)\Psi\left(x\right)$, which
upon substitution into Eq.$\ $\eqref{eq:1dhetro} and its multiplication
by $\rho^{-1/2}\left(x\right)$ provides a different representation
of the Hamiltonian, namely, 
\begin{equation}
\hat{H}=-\frac{1}{\sqrt{\rho\left(x\right)}}\ddx\mu\left(x\right)\ddx\frac{1}{\sqrt{\rho\left(x\right)}}.\label{eq:hamilv3}
\end{equation}
For real moduli, this Hamiltonian is Hermitian if it operates on functions
that vanish at the boundary of the problem domain, and therefore the
eigenfunctions $\left\{ \Psi\left(x\right)\right\} $ are orthogonal
one to each other; the application of NHQM perturbation theory to
such Hermitian systems in 1D elastodynamics is demonstrated first.

\section{\label{sec:EP}NHQM Perturbation theory for elastodynamics: the model
problem of a finite slab}

In quantum mechanics, the standard time-independent Rayleigh-Schrödinger
theory provides the solutions of a perturbed Hamiltonian in terms
of a series expansion about an Hermitian Hamiltonian. The NHQM formalism
determines the radius of convergence by extending the perturbation
to the complex plane and calculating the EP---the point at which
the perturbed Hamiltonian has a non-Hermitian degeneracy \citep{moiseyev2011book,Miri2019science,Ozdemir2019cr}.
The process is exemplified in this Section, by calculating first the
eigenfrequencies and eigenmodes of an Hermitian Hamiltonian that models
an elastodynamic system made of purely elastic and finite slabs; subsequently,
we determine the convergence radius of the elastodynamic Rayleigh-Schrödinger
expansion by calculating the EP in the perturbed non-Hermitian Hamiltonian
spectrum. 

Thus, we truncate the assembly at $x=\pm L$, and fix the boundaries
such that the displacement field vanishes at the edges. The standard
procedure to calculate the real frequencies starts with the ansatz
\begin{equation}
U^{(0)}\left(x\right)=\begin{cases}
A\cos\phaseone kx+B\sin\phaseone kx, & x<|\period|,\\
C_{+}\sin{\phasetwo k(x-L)}, & l<x<L,\\
C_{-}\sin{\phasetwo k(x+L)}, & -L<x<-\period,
\end{cases}\label{eq:usoltwophasefinite}
\end{equation}
where owing to Eq.$\ $\eqref{eq:eom shear pressure} and the continuity
of $u(x,t)$ 
\begin{equation}
\omega^{2}=c_{i}^{2}k_{i}^{2},\quad i=a,b,\label{eq:nodispersion}
\end{equation}
and hence $\phaseone k$ and $\phasetwo k$ are related via
\begin{equation}
\frac{\phaseone k}{\phasetwo k}=\frac{\phasetwo c}{\phaseone c}.\label{eq:snell}
\end{equation}
The continuity of the spatial parts of the displacement and stress
at $x=\pm\period$ takes the form 
\begin{align}
A\cos\phaseone k\period & =C\sin{\phasetwo k(\period-L)},\label{eq:dispcontfinite}\\
-A\phaseone{\mu}\phaseone k\sin\phaseone k\period & =\phasetwo{\mu}\phasetwo kC\cos{\phasetwo k(\period-L)},\label{eq:tractioncontfinite}
\end{align}
from which the relation between the amplitudes $A$ and $C$ is determined.
The resultant transcendental equation for the eigenfrequencies is
\begin{align}
 & \tan\frac{\omega\period}{\phaseone c}=\impedancemis\,\tan\frac{\omega(\period-L)}{\phasetwo c},\quad & (\mathrm{odd}\:\mathrm{modes})\label{eq:oddfinite}\\
 & \cot\frac{\omega\period}{\phaseone c}=-\impedancemis\,\tan\frac{\omega(\period-L)}{\phasetwo c},\quad & (\mathrm{even}\:\mathrm{modes}).\label{eq:evenfinite}
\end{align}

We denote the eigenfrequencies by $\{\omega_{m}^{(0)}\}_{m\in\mathbb{N}}$,
and the corresponding transformed eigenfunctions of Eq.$\ $\eqref{eq:hamilv3}
by $\Psi_{m}^{(0)}(x)$.  It is clear that $\{\omega_{m}^{(0)}\}_{m\in\mathbb{N}}$
are real and the Hamiltonian is indeed Hermitian.

Consider next another assembly, obtained by replacing the right half
of the central slab by a slab whose shear stiffness is $\alpha$.
The resultant Hamiltonian can be written as a sum of the Hamiltonian
of the original medium, denoted $\hat{H}^{(0)}$, and a perturbation
$\parameter\hat{H}^{(1)}$, where 
\begin{equation}
\hat{H}^{(1)}=-\frac{1}{\sqrt{\rho_{a}}}\ddx\ddx\frac{1}{\sqrt{\rho_{a}}}\label{eq:perturbation}
\end{equation}
operates on functions over $0<x<l$. Up to a critical value of $\parameter$,
NHQM perturbation theory can deliver the response of the perturbed
assembly, in terms of $\Psi_{m}^{(0)}(x)$ as the zero-order solutions
\citep{cohen1991quantum}. Using the standard time-independent Rayleigh-Schrödinger
theory, we obtain the $n^{\mathrm{th}}$ order correction terms $\omega^{(j)}$
and $\Psi_{m}^{(j)}(x)$, namely, 
\begin{equation}
\omega_{m}^{2}(x;\parameter)=\sum_{j=0}^{\infty}\parameter^{j}\omega_{m}^{2(j)},\quad\Psi_{m}(x;\parameter)=\sum_{j=0}^{\infty}\parameter^{j}\Psi_{m}^{(j)};
\end{equation}
the convergence of these sums is limited to values of $\parameter$---including
complex values---inside a circle in the complex plane whose origin
is 0 and its radius is denoted $|\lep|$. This radius equals the radius
of the complex branch point at which two adjacent modes coalesce.
Thus, a non-Hermitian degeneracy is obtained when the conditions 
\begin{equation}
U_{m}(\lep)=U_{m\pm1}(\lep)\equiv U_{m}^{EP},\label{condI}
\end{equation}
and 
\begin{equation}
\Psi_{m}(x;\lep)=\Psi_{m\pm1}(x;\lep)\equiv\Psi_{m}^{EP}(x)\label{condII}
\end{equation}
are satisfied. Since for any value of $\parameter\ne\lep$ the two
modes $\Psi_{m}(x;\lep)$ and $\Psi_{m\pm1}(x;\lep)$ are orthogonal
one to another, at the EP $\Psi_{m}^{EP}(x)$ is self-orthogonal,
as the two solutions coalesce \citep{moiseyev2011book}. To determine
$\lep$, we first represent $\hat{H}^{(0)}$ and $\hat{H}^{(1)}$
using the matrices $\hmatz$ and $\hmato$, defined by 
\begin{equation}
\mathsf{H}^{(\alpha)}{}_{mn}=\int_{\mathscr{I}}\hat{\Psi}_{m}^{(0)}\hat{H}^{(i)}\hat{\Psi}_{n}^{(0)}\mathrm{d}x,\quad i=0,1,\label{eq:hcomponent}
\end{equation}
where $\mathscr{I}=\left[-L,L\right]$ and $\left[0,l\right]$ when
$i=0$ and 1, respectively, and 
\begin{equation}
\hat{\Psi}_{m}^{(0)}=\Psi_{m}^{(0)}/\int_{-L}^{L}\Psi_{m}^{2(0)}\mathrm{d}x.\label{eq:normalization}
\end{equation}
Note that the standard procedure to derive orthogonality relations
for real functions provides 
\begin{equation}
\left\langle \hat{\Psi}_{m}^{(0)},\hat{\Psi}_{n}^{(0)}\right\rangle \coloneqq\int_{-L}^{L}\hat{\Psi}_{m}^{(0)}\hat{\Psi}_{n}^{(0)}\mathrm{d}x=\delta_{mn}.\label{eq:psiorth}
\end{equation}
 We are now at the position to seek the smallest $\parameter$ for
which the matrix 
\begin{equation}
\hp(\parameter)=\hmatz+\parameter\hmato,\label{eq:perturbed H}
\end{equation}
has an eigenvalue multiplicity, using a modified Newton's method \citep{Mailybaev2006}.
To proceed with numerical computations, we consider by way of example
a middle slab made of PMMA, which is bonded between two steel slabs,
whose properties are
\begin{equation}
\begin{array}{r@{}lc}
 & \phaseone{\rho}=1200\,\mathrm{kg\,m^{-3}},\quad\phaseone{\mu}=1.21\,\mathrm{GPa}, & l=1\,\mathrm{cm},\\
 & \phasetwo{\rho}=7800\,\mathrm{kg\,m^{-3}},\quad\phasetwo{\mu}=78.85\,\mathrm{GPa}, & L=3\,\mathrm{cm}.
\end{array}\label{eq:velocities-1}
\end{equation}
For simplicity, we truncate the size of $\mathsf{H}^{(\alpha)}$ to
$2\times2$ using the first odd and even modes. 

\begin{figure}[t]
\centering\sidesubfloat[]{\label{fig:inversew}\includegraphics[width=0.65\textwidth]{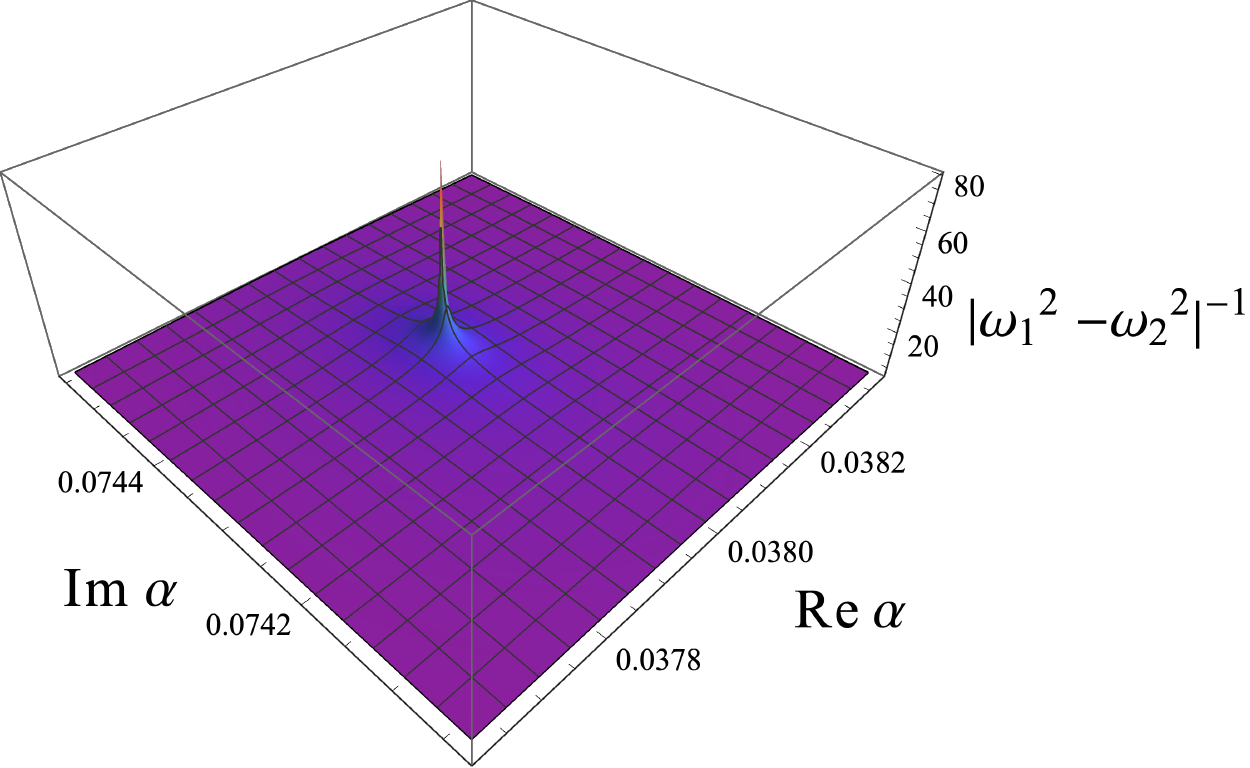}} \quad\\ \sidesubfloat[]	{\label{fig:selforthogonal}\includegraphics[width=0.65\textwidth]{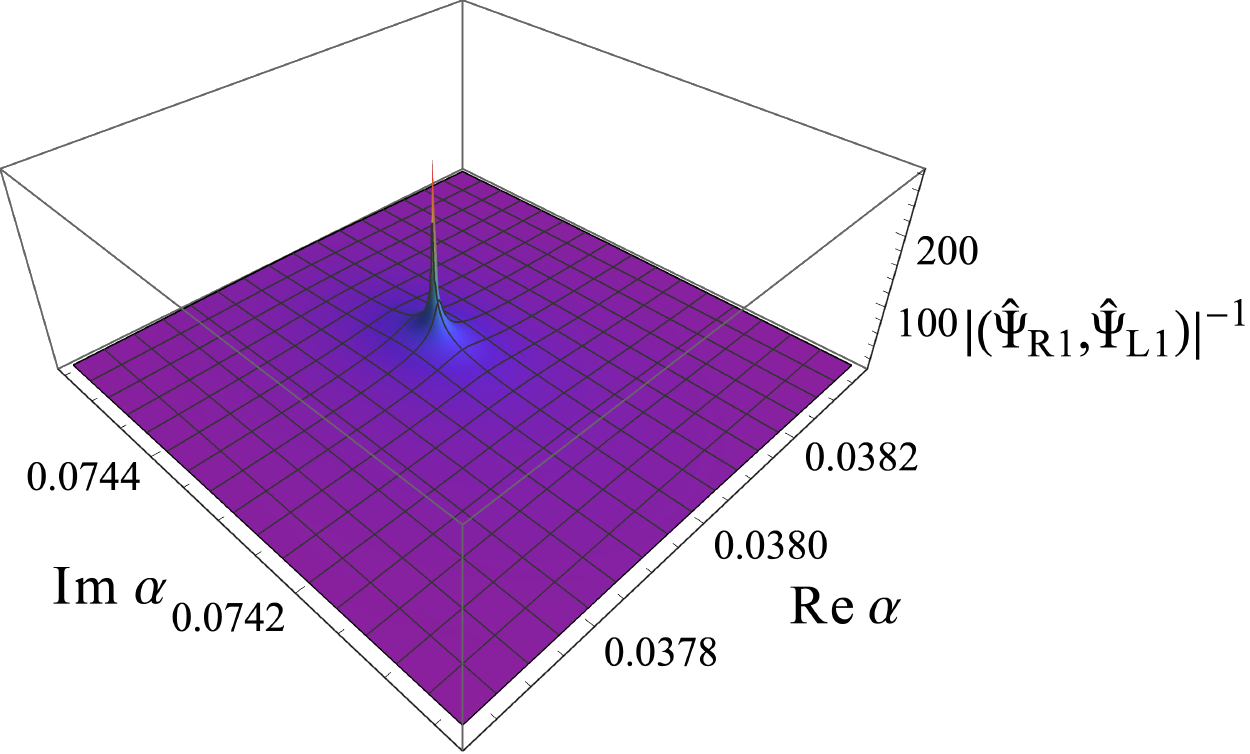}} \quad\\ \sidesubfloat[]	{\label{fig:riemannre}\includegraphics[width=0.65\textwidth]{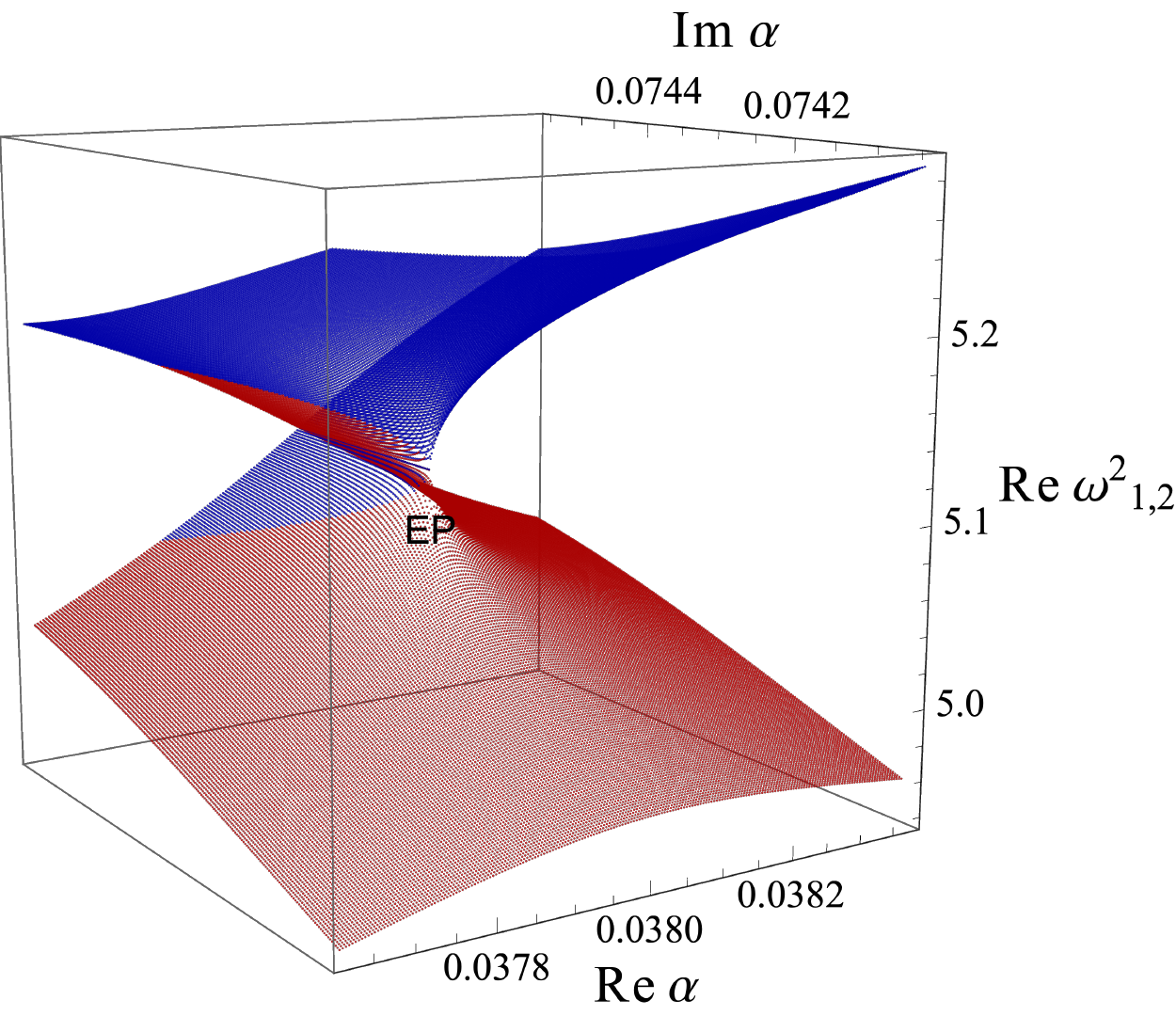}}\quad\\ \sidesubfloat[]	{\label{fig:riemannim}\includegraphics[width=0.65\textwidth]{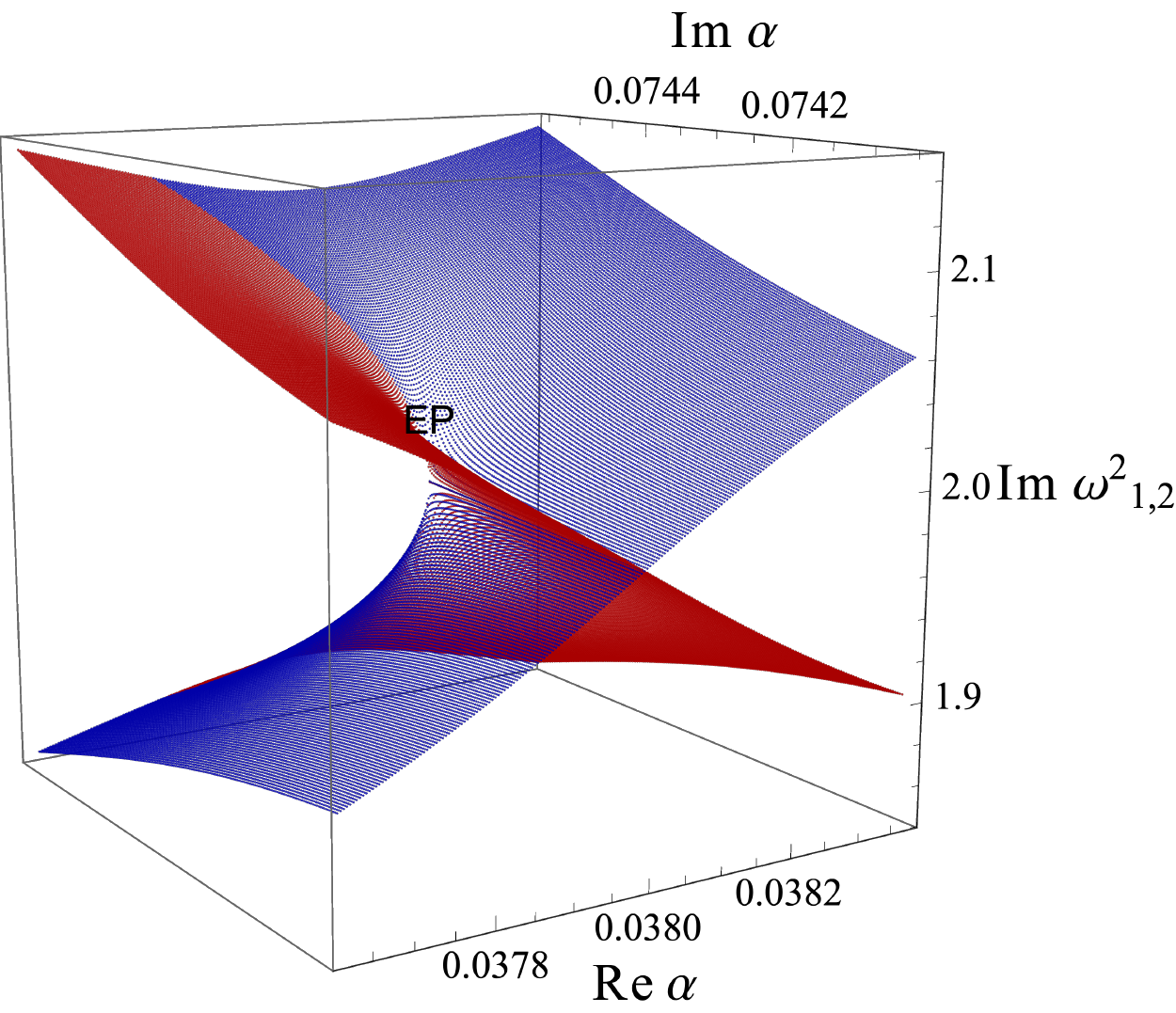}}\caption{(a) The inverse of the absolute value of difference between the  eigenvalues as function of $\parameter$. Observe that $\lim_{\parameter\rightarrow\lep}\left|\omega_{1}^{2}-\omega_{2}^{2}\right|^{-1}=\infty$. (b) The inverse of the inner product between the first right and left  eigenvectors as function of $\parameter$. Observe that $\lim_{\parameter\rightarrow\lep}\left|\left\langle \hat{\Psi}_{\mathrm{R}1},\hat{\Psi}_{\mathrm{L}1}\right\rangle \right|^{-1}=\infty$. (c)
The real part of the eigenvalues as function of $\protect\parameter$.
The red and blue surfaces correspond to $\omega_{1}^{2}$ and $\omega_{2}^{2}$,
respectively. (d) The imaginary part of the eigenvalues as function
of $\protect\parameter$. (Same color legend as in the previous panel.)
At $\alpha=\protect\lep$ the two eigenvalues coalesce to the complex
$\omega^{EP}$. }

{\small{}{}\label{fig:ep}}{\small\par}
\end{figure}

The results are shown in Fig.$\ $\ref{fig:ep} in a dimensionless
form (lengths are divided by $L$, mass densities and shear moduli
are divided by the mean value of the quantity when averaged between
the two phases). \floatsetup[figure]{style=plain,subcapbesideposition=top}

Specifically, Fig.$\ $\ref{fig:inversew} shows the inverse of the
absolute value of the difference between the two eigenvalues of $\hp$
versus Re$\,\parameter$ and Im$\,\parameter$; the peak at $\alpha\simeq0.38+0.074i$
identifies $\lep$, and hence the radius of convergence. We note that
such a value of $\alpha$, i.e., with a positive imaginary part, physically
corresponds to a viscoelastic slab, and hence realzing this EP does
not require any gain.  Fig.$\ $\ref{fig:selforthogonal} shows the
inverse of the inner product between the first right ($\hat{\Psi}_{\mathrm{R}1}$)
and left ($\hat{\Psi}_{\mathrm{L}1}$) eigenvectors as function of
$\parameter$. Indeed, we observe that $\lim_{\alpha\rightarrow\lep}\left|\left\langle \hat{\Psi}_{\mathrm{R}1},\hat{\Psi}_{\mathrm{L}1}\right\rangle \right|^{-1}=\infty$,
confirming that this is a non-Hermitian degeneracy, as that the corresponding
functions are self-orthogonal \footnote{For asymmetric matrices, the notion of orthogonality is replaced with
bi-orthogonality of right and left eigenvectors \citep{moiseyev2011book}}. Finally, we show that the spectrum in the vicinity of the EP exhibits
a Riemann surface structure---the signature of non-Hermitian degeneracy---by
plotting the real (Fig.$\ $\ref{fig:riemannre}) and imaginary (Fig.$\ $\ref{fig:riemannim})
parts of $\omega_{1}^{2}$ (red surface) and $\omega_{2}^{2}$ (blue
surface). We denote for later use the frequency at the EP by $\omega^{EP}$,
such that 
\begin{equation}
\omega^{EP}\coloneqq\omega_{1}\left(\alpha=\lep\right)=\omega_{2}\left(\alpha=\lep\right).\label{eq:wep}
\end{equation}


\begin{center}
\par\end{center}

\section{\label{sec:Application-to-mass-sensing}Application to mass sensing}

\begin{figure}[t]
\centering\sidesubfloat[]{\label{fig:splitting}\includegraphics[width=0.8\textwidth]{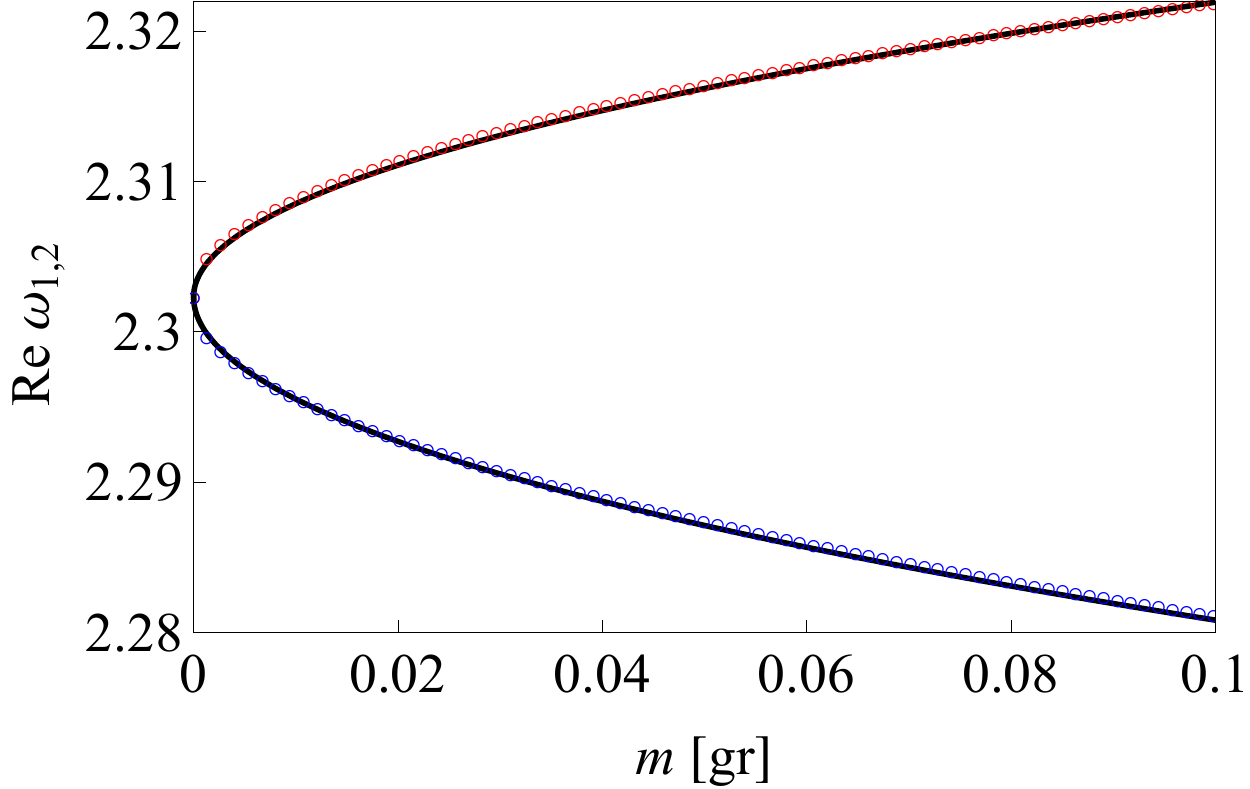}}\\\sidesubfloat[]	{\label{fig:gradient}\includegraphics[width=0.8\textwidth]{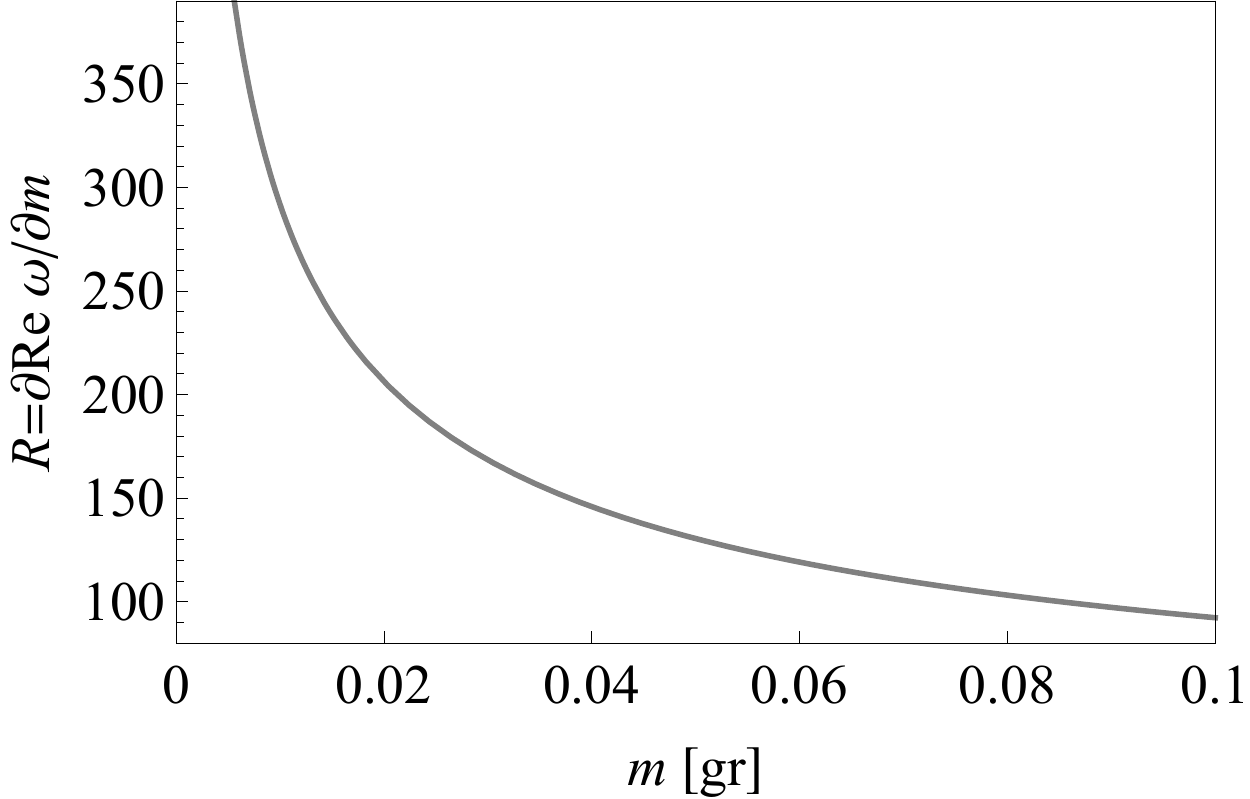}}\caption{(a) The real part of the first (blue circles) and second (red circles)
eigenfrequencies of the our elastodynamic system, when operating at
an EP and augmented by a deposited mass $m$ (in grams). The solid
lines are the functions \eqref{eq:scaling}, which are proportional
to $\sqrt{m}$, and their agreement with the calculated frequencies
confirms the square-root nature of the frequency splitting. By contrast,
conventional mechanical sensors are linear in $m$. (b) The mass responsivity
$R=\partial\mathrm{Re}\,\omega/\partial m$ of device as function
of $m$. Notice the theoretical infinite responsivity in the limit
of an infinitesimal mass.}

{\small{}{}\label{fig:sensor}}{\small\par}
\end{figure}

We demonstrate next how the unique topology near the EP can be harnessed
to design a mass sensor with enhanced sensitivity, based on the aforementioned
assembly when  tuned to operate at the EP. First, we recall that
standard mechanical mass sensors are based on the shift in the mechanical
resonant frequency, owing to any deposited mass \citep{Liu2013qf}.
For small masses, the shift is linear in the perturbation \citep{Boisen2011hc,He2015rp}.
This linear relation can be interpreted as the first term in the Taylor
series of the frequency as function of the mass
\begin{equation}
\mathrm{Re}\,\omega\left(m\right)-\mathrm{Re}\,\omega\left(0\right)=S_{T}m+\mathcal{O}\left(m^{2}\right),\label{eq:Taylor}
\end{equation}
where $m$ is the mass of the deposited element, $\mathrm{Re}\,\omega\left(0\right)$
is the (real part of the) resonant frequency of the unperturbed system,
and $S_{T}=\partial\mathrm{Re}\,\omega\left(m\right)/\partial m$
at $m=0$. By contrast, there is not a Taylor expansion of the frequency
shift from an EP, however it does admit a Puiseux Series. In case
when the EP is of two eigenvalues and eigenmodes, we have that 
\begin{equation}
\mathrm{Re}\,\omega\left(m\right)-\mathrm{Re}\,\omega^{EP}=S_{P}\sqrt{m}+\mathcal{O}\left(m\right),\label{eq:Puiseux}
\end{equation}
with some coefficient $S_{P}$. Sensing of a device is thus quantified
by the so-called the mass responsivity $R=\partial\mathrm{Re}\,\omega/\partial m$
\citep{Ekinci2004hb,He2015rp}. It is clear that conventional sensors
have a finite $R$ as $m\rightarrow0$, whereas for EP-based sensors
\begin{equation}
\lim_{m\rightarrow0}\frac{\partial\mathrm{Re}\,\omega\left(m\right)}{\partial m}\wasypropto\lim_{m\rightarrow0}\frac{1}{\sqrt{m}}=\infty,\label{eq:sensitivity}
\end{equation}
i.e., theoretically an infinite sensitivity, which in practice is
limited by the resolution of the frequency measurement. This feature
has been employed for sensors in different physical systems \citep{Wiersig2014prl,Wiersig2016pra},
and specifically in systems comprising optomechanical cavities coupled
by mechanical resonators \citep{Djorwe2019prapplied}. Here, we apply
this approach to the elastodynamic system described in the previous
section, emphasizing that its EP does not require any realization
of gain, as that assembly comprises only elastic and viscoelastic
slabs.  This is carried out by calculating the eigenfrequencies when
a discrete element with mass $m$ is deposited at the center of the
assembly whose Hamiltonian is given by Eqs.$\ $\eqref{eq:perturbed H}-\eqref{eq:velocities-1}
with $\alpha=\lep$. The mass is modeled by replacing $\rho\left(x\right)$
with $\rho\left(x\right)+\delta\rho\left(x\right)$ in the Hamiltonian
\eqref{eq:hamilv3} where $\delta\rho\left(x\right)=\delta\rho_{0}$
over $-L/50<x<L/50$ such that $m=\delta\rho_{0}L/100$ \footnote{This choice has been made for numerical convenience, and approximates
the model $\delta\rho\left(x\right)=m\delta\left(x\right)$, where
$\delta\left(x\right)$ is the Dirac delta, such that its integral
over $x$ equals $m$. }, and calculating the resultant eigenvalues of Eq.$\ $\eqref{eq:perturbed H}.
Fig.~\ref{fig:splitting} depicts (the real part of) the first (blue
circles) and second (red circles) eigenfrequencies for representative
values of the deposited mass (in grams). It is shown how frequency
splitting occurs owing to the added mass, in a manner that is nonlinear
in the perturbation---the smaller the mass, the greater the relative
change. The solid lines are the functions 
\begin{equation}
\mathrm{Re}\,\omega^{EP}+S_{P}\sqrt{m},\label{eq:scaling}
\end{equation}
where $S_{P}=-2.02$ and $1.845$ for the lower and upper curves,
respectively; the matching between Eq.$\ $\eqref{eq:scaling} and
the calculated frequencies thereby confirms the conjectured square
root nature of the frequencies dependency in the deposited mass. Again,
we emphasize that the scaling is linear in conventional mechanical
sensors based on the shift of the resonant frequency, therefore inferior
for extremely small masses. 

The enhanced sensitivity near the EP is further highlighted in Fig.~\ref{fig:gradient},
by plotting the mass responsivity as function of $m$, using the derivative
of the fitted function \eqref{eq:scaling} for the higher frequency.
Thereby, we show the theoretical infinite responsivity in the limit
of an infinitesimal mass. 

\floatsetup[figure]{style=plain,subcapbesideposition=top}

\section{\label{sec:radiation}A Non-Hermitian Model problem with outgoing
boundary conditions }

 We pursue next the analogy drawn in Section \ref{sec:Similarities-and-differences}
between the infinite elastic assembly and the model problem in NHQM
of a particle in a box with outgoing boundary conditions, thereby
presenting the physical interpretation of this theory for elastodynamics.
Specifically, we will demonstrate that the NHQM approach will provide
the so-called leaky eigenmodes of the system, whose imaginary part
of the eigenvalues delivers the decay rate of the elastic energy in
the middle slab. 

\begin{figure}[h] 
\includegraphics[width=1\textwidth]{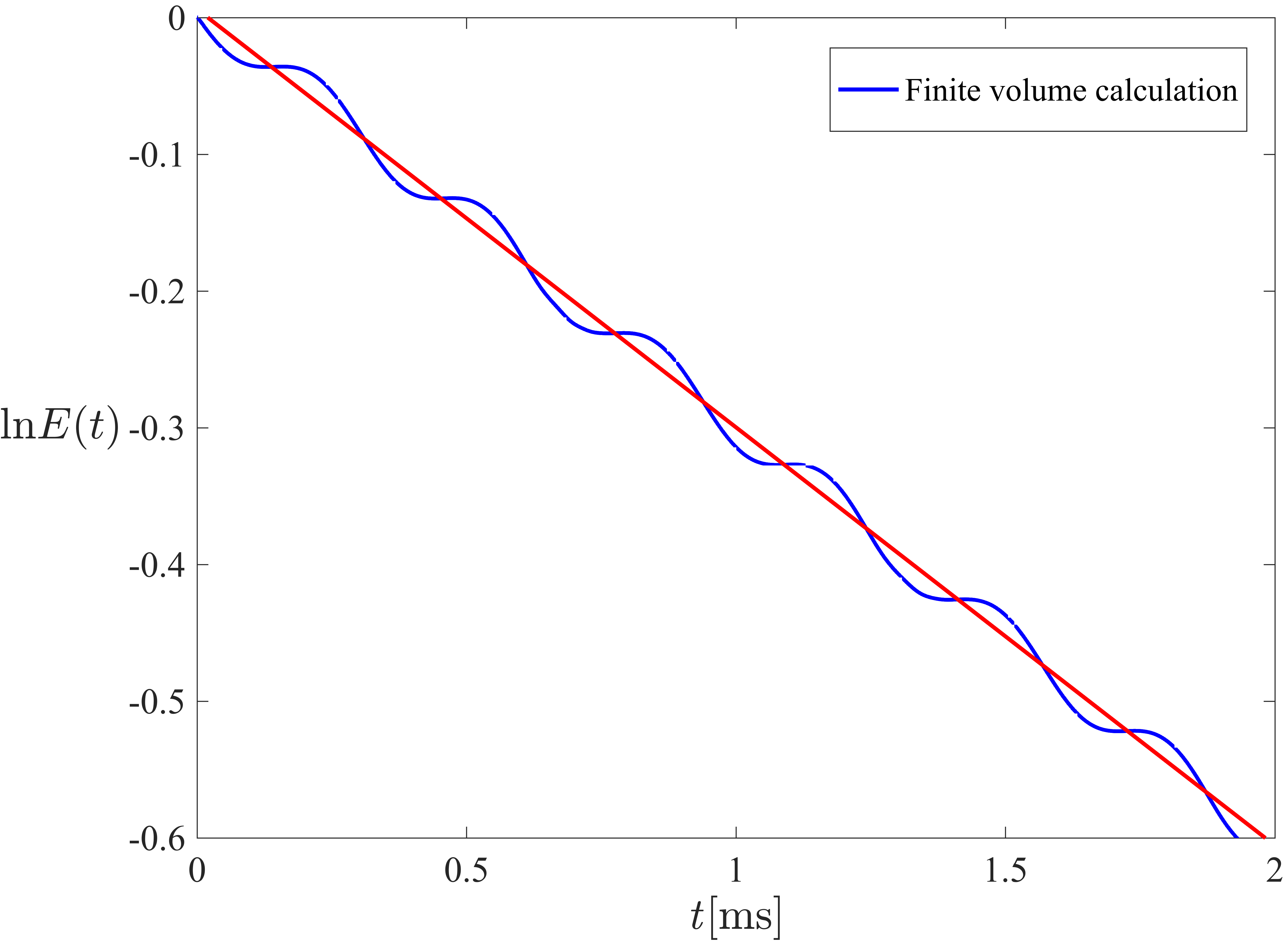} \caption{\label{fig:energyrate}Log of the mechanical energy  stored in the PMMA  slab (normalized by as function of $t\,$(ms), calculated using a finite-volume scheme (blue curve). The slope of its linear interpolation (red curve) matches -2Im$\,\omega$. } \end{figure}

Accordingly, when the steel slabs now extend to $\pm\infty$ and the
PMMA slab is released from some arbitrary initial shear deformation,
we impose outgoing boundary conditions and seek solutions to $\ux$
in the form
\begin{equation}
\ux=\begin{cases}
A\cos\phaseone kx+B\sin\phaseone kx, & x<|\period|,\\
C_{+}e^{i\phasetwo k(x-l)}, & x>\period,\\
C_{-}e^{-i\phasetwo k(x+l)}, & x<-\period.
\end{cases}\label{eq:usoltwophase-1}
\end{equation}
(The relation between $k_{a},k_{b}$ and $\omega$ remains as in the
Hermitian problem.) 
The continuity of the displacement and stress at $x=\pm\period$ now
reads 
\begin{align}
A\cos\phaseone k\period\pm B\sin\phaseone k\period & =C_{\pm},\label{eq:dispcont-1}\\
\phaseone{\mu}\phaseone k(\mp A\sin\phaseone k\period+B\cos\phaseone k\period) & =\pm C_{\pm}\phasetwo{\mu}i\phasetwo k.
\label{eq:tractioncont-1}
\end{align}
Manipulating these equations provides 
\begin{align}
 & i\tan\phaseone k\period=\impedancemis,\quad & (\mathrm{odd}\:\mathrm{modes})\label{eq:odd}\\
 & i\cot\phaseone k\period=-\impedancemis,\quad & (\mathrm{even}\:\mathrm{modes})\label{eq:even}
\end{align}
where the impedance mismatch $\impedancemis$ is 
\begin{equation}
\impedancemis=\frac{\phaseone{\mu}\phaseone k}{\phasetwo{\mu}\phasetwo k}=\frac{\phaseone{\mu}\phasetwo c}{\phasetwo{\mu}\phaseone c}=\sqrt{\frac{\phaseone{\mu}\phaseone{\rho}}{\phasetwo{\mu}\phasetwo{\rho}}}.
\end{equation}
Eqs.$\ $\eqref{eq:odd}-\eqref{eq:even} are solved by 
\begin{eqnarray}
 & \phaseone k\period & =-i\,\mathrm{arctanh}\impedancemis+\frac{m\pi}{2},\:\:m\in\mathbb{Z},\label{eq:dispersioninf}
\end{eqnarray}
where odd and even $m$ correspond to even and odd modes, respectively.
Hence, there are infinitely many discrete complex solutions with a
different real part and the same imaginary part; there are no bound
states associated with real solutions. The obtained roots are the
poles of the scattering matrix, which only in the non-Hermitian formalism
of quantum mechanics delivers fundamental information on the modes,
without the need to carry out wave packet calculations. For example,
in NHQM the imaginary part of the poles provides the resonance width,
or rate of decay of resonance state \citep{moiseyev2011book}. Here,
by analogy, the imaginary part in Eq.$\ $\eqref{eq:dispersioninf}
should provide information on the decay rate of the mechanical energy.
This relation is demonstrated in Fig.$\ $\ref{fig:energyrate}, where
the log of the mechanical energy stored in the PMMA slab 
\begin{equation}
E_{a}(t)=\frac{1}{2}\int_{-\period}^{\period}\left(\phaseone{\mu}u_{,x}^{2}+\phaseone{\rho}u_{,t}^{2}\right)\mathrm{d}x
\end{equation}
is evaluated as function of $t$, when calculated using the finite
volume method \citep{ziv2019b} for some (real) arbitrary initial
conditions. Indeed, the slope of its linear interpolation (red curve)
matches -2Im$\,\omega$, and is independent of the form of the initial
conditions. 

\section{\label{sec:theory-for-real}theory for real perturbations in non-hermitian
elastodynamics: the 1D model problem}

In this last part, we are interested in developing a perturbation
theory to the latter problem, noting that the obstacle lies in the
divergence of $\Psi(x)$ at $\pm\infty$. Therefore, the orthogonality
relations \eqref{eq:hcomponent} no longer hold, and the components
of $\hmatz$ are unbounded when calculated according to Eq.$\ $\eqref{eq:psiorth}.
To overcome these obstacles,  we first apply the complex scaling
transformation $\left(x-l\right)\rightarrow\left(x-l\right)e^{i\theta}$
for $\Psi(x>l)$, with sufficiently large and real $\theta$ \footnote{See Chapt.$\ $5 in Ref.$\ $\citep{moiseyev2011book} and the references
therein}. In these rotated coordinates, the transformed function \begin{equation}
\begin{aligned}
&\Psi_{n}\left(x>l;\theta\right)=e^{ik_{bn}\left(x-l\right)e^{i\left(\theta-\fn\right)}}\\
&=e^{i\left|k_{bn}\right|\left(x-l\right)\cos\left(\theta-\fn\right)-\sin\left(\theta-\fn\right)}e^{-\left|k_{bn}\right|\left(x-l\right)\sin\left(\theta-\fn\right)}
\end{aligned}
\end{equation} with $\text{\ensuremath{\tan\fn}}=-\left(n\pi\right)^{-1}\mathrm{arctanh}\gamma$,
vanishes at infinity, owing to the second decaying exponent. Similarly,
$\Psi(x<-l;\theta)$ vanishes at $-\infty$ by applying the transformation
$\left(x+l\right)\rightarrow\left(x+l\right)e^{i\theta}$. 

To establish next an orthonormal basis set, we replace the scalar
product of the Hermitian formalism with the NHQM \emph{c-product}
\citep{moiseyev2011book}, namely, \begin{equation}
\begin{aligned}
&\innerc{\Psi_{n}}{\Psi_{n}}\coloneqq\int_{-\infty}^{-l}\Psi_{n}^{2}\left(x;\theta\right)\mathrm{d}x+\int_{-l}^{l}\Psi_{n}^{2}\left(x\right)\mathrm{d}x+\int_{l}^{\infty}\Psi_{n}^{2}\left(x;\theta\right)\mathrm{d}x,\\
\end{aligned}
\label{eq:norm}
\end{equation} To show that this product indeed delivers such a set, consider first
the third term in Eq.$\ $\eqref{eq:norm}.   Since the scaled function
vanishes at infinity, the integral is zero at its upper limit, and
we are left with its value at $x=l$, such that 
\begin{align}
\int_{l}^{\infty}\rho_{b}C_{n+}^{2}e^{2ik_{bn}\left(x-l\right)e^{i\theta}}e^{i\theta}\mathrm{d}x=\frac{i\rho_{b}}{4k_{bn}}\left(1\pm\cos2k_{an}l\right),\label{eq:intext}
\end{align}
where we used that fact that $C_{n+}=\cos k_{an}l$ for the even
modes (with a plus sign inside the brackets), and $C_{n+}=\sin k_{an}l$
for the odd modes (minus sign). Owing to symmetry, this is also the
value of the first integral in Eq.$\:$\eqref{eq:norm}, and the remaining
integral amounts to
\begin{equation}
\int_{-l}^{l}\Psi_{n}^{2}\left(x\right)\mathrm{d}x=\rho_{a}\left(l\pm\frac{\sin2k_{an}l}{2k_{an}}\right),\label{eq:integrationl}
\end{equation}
where the plus and minus signs correspond to even and odd modes, respectively.
We can now redefine the basis \eqref{eq:normalization} to 
\begin{equation}
\hat{\Psi}_{n}^{(0)}=\Psi_{n}^{(0)}/\innerc{\Psi_{n}}{\Psi_{n}},\label{eq:normalization2}
\end{equation}
and observe that 
\begin{equation}
\innerc{\Psi_{n}\left(x\right)}{\Psi_{m}\left(x\right)}=\delta_{nm},\label{eq:orthnh}
\end{equation}
when invoking Eq.$\ $\eqref{eq:dispersioninf}.

As in Section \ref{sec:EP}, we replace again a part of the middle
slab by a third constituent whose shear stiffness is $\parameter$;
now, however, we leave the location, say $d$, and length, say $D-d$,
of the replacement as parameters. Using the framework developed in
this Section, we can derive the eigenstates of the perturbed Hamiltonian
as an expansion about a non-Hermitian Hamiltonian with real parameters.
Specifically, we have that 

\begin{equation}
\mathsf{H}\left(\alpha,d,D\right)=\mathsf{H}^{\left(0\right)}+\alpha\mathsf{H}^{\left(1\right)}\left(d,D\right),\label{eq:perturbedinfinite}
\end{equation}
where $\mathsf{H}^{\left(0\right)}$ is diagonal with complex eigenvalues
associated with Eq.$\ $\eqref{eq:dispersioninf}, and \begin{equation}
\begin{aligned}
&\mathsf{H}^{\left(1\right)}_{nm}=\int_{D}^{d}\Psi_{n}\left(x\right)\ddxsa{}\Psi_{m}\left(x\right)\mathrm{d}x=k_{am}^{2}\rho_{a}\int_{d}^{D}\Psi_{n}\left(x\right)\Psi_{m}\left(x\right)\mathrm{d}x,
\end{aligned}
\label{eq:h1compinf}
\end{equation}is a complex asymmetric matrix $\mathsf{\hat{H}}^{\left(1\right)}$
that depends nonlinearly on $d$ and $D$. (The resultant closed-form
expressions are omitted here, for brevity). Notably, our developments
further establish a platform for constructing degeneracies by real
perturbations. 

\section{\label{sec:Summary}Summary and outlook}

Motivated by the development of non-Hermitian quantum mechanics and
the transfer of concepts from quantum theories to the macroscopic
scale, we here revisit the connection between the time-independent
Schrödinger equation and the one-dimensional elastodynamics equation.
Using a simple transformation, we have first identified the term that
appears in the elastodynamics equation and absent from the quantum
mechanics equation. This derivation allowed us to determine the condition
under which the two equations are equivalent. 

Subsequently, we have showed the physical interpretation and application
of different tools from non-Hermitian quantum mechanics in elastodynamics,
including the time-independent Rayleigh-Schrödinger perturbation theory
to calculate the dynamic response of a finite elastic assembly; the
non-Hermitian formalism of this theory to determine the perturbation
series radius of convergence and exceptional point in the spectrum
of the assembly; calculation of leaky modes and energy decay in an
open elastic assembly using the poles of the scattering matrix; and
the complex scaling transformation for establishing a basis from the
corresponding divergent eigenfunctions. 

Importantly, we have introduced a framework to analyze and design
non-Hermitian degeneracies by real perturbations. These degeneracies
have great potential in applications such as ultra-sensitive sensors
and unidirectional energy scatterers, for which our approach offers
a way to access without the need for gain or $\pt$ symmetry. As a
concrete application, we have demonstrated how an elastic slab assembly
can function as enhanced mass sensor, when designed according to our
analysis to exhibit two degenerate shear states. Using algebraic arguments
and numerical calculations, we showed in Section \ref{sec:Application-to-mass-sensing}
how the mass responsivity of this sensor surpasses the responsivity
of conventional mechanical sensors owing to the square-root topology
near the exceptional point in its spectrum. 

We expect that our simplified demonstration of the potential that
the tools non-Hermitian quantum mechanics has in elastodynamics will
pave the way for further developments in more complex, practical elastic
systems. Examples include periodic composites and homogenization \citep{milton2002theory,Antonakakis2014JMPS,torrent2014prb,Shmuel2016JMPS,LUSTIG2018jmps,Mokhtari2019arxiv},
anisotropic media \citep{stroh1962JMP,ting1996anisotropic}, and elastodynamics
in higher dimensions \citep{elasticity198}.
\begin{acknowledgments}
This research was supported in parts by the I-Core: the Israeli Excellence
Center \textquotedbl Circle of Light\textquotedbl , the Israel Science
Foundation (grants No.$\ $1530/15 and 1912/15), the United States-Israel
Binational Science Foundation (grant No.$\ $2014358), and the ministry
of science and technology. We thank Ron Ziv for sharing his MatLab
code, and anonymous reviewers for constructive comments that helped
us improve this paper. 
\end{acknowledgments}

\bibliographystyle{unsrt}
\bibliography{bibtexfiletot}

\end{document}